\documentclass{ifacconf}

\usepackage{graphicx}      
\graphicspath{{./graphics/}}
\usepackage{natbib}        

\usepackage{xcolor}
\usepackage{amsmath,amssymb,amsfonts}
\usepackage{dsfont}
\usepackage{enumitem}

\newcommand{\tr}[1]{\tilde{#1}}

\newcommand{\norm}[1]{\left\lVert#1\right\rVert}

\newcommand{\ONE}{\mathbf{1}}
\newcommand{\ZERO}{\mathbf{0}}

\newcommand{\s}[2][]{^{#2^{#1}}}
\newcommand{\tbf}[2][]{\mathbf{#1\tilde{#2}}}
\newcommand{\rev}[1]{{#1}}

\usepackage{psfrag}
\usepackage{subfigure}

\begin{document}
\begin{frontmatter}

\title{Distributed MPC with Prediction of Time-Varying Communication Delay\thanksref{footnoteinfo}} 

\thanks[footnoteinfo]{Partial financial support by the DFG within the Priority Program \textit{Cyber-Physical Networking} (SPP 1914) is gratefully acknowledged.}

\author[UKS]{J. Hahn}
\author[FU]{R. Schoeffauer}
\author[FU]{G. Wunder} 
\author[UKS]{O. Stursberg} 

\address[FU]{Communication and Information Theory, Freie Universit\"at Berlin,\\ \{r.schoeffauer,  g.wunder\}@fu-berlin.de}
\address[UKS]{Control and System Theory, Universit\"at Kassel,\\ \{jhahn, stursberg\}@uni-kassel.de}

\begin{abstract}                
The novel idea presented in this paper is to interweave distributed model predictive control with a reliable scheduling of the information that is interchanged between local controllers of the plant subsystems. To this end, a dynamic model of the communication network and a predictive scheduling algorithm are proposed, the latter providing predictions of the delay between sending and receiving information. These predictions can be used by the local subsystem controllers to improve their control performance, as exemplary shown for a platooning example.
\end{abstract}


\end{frontmatter}

\section{Introduction}
Two major trends can be recognized in the modern information society: one is that more and more physical systems used on a daily basis are equipped with digital controllers, sensors, and actuators (leading to \textit{embedded systems}) -- secondly, these embedded systems are connected to a global information network (the \textit{cyber space}). Bringing these two trends together is a current main challenge in engineering. Systems in which communication and control are formulated within a common mathematical model are called \textit{Cyber-Physical-Systems} (CPS), see \citep{lee_general_cps}. In CPS, the traditional modeling of plant and controller is extended by a model of the (wireless) communication between actuators, sensors, and multiple control units. In the standard setting, the overall objective remains in the realm of control, while stringent requirements are formulated for the communication. Of crucial importance is the latency in the wireless network, i.e., the delay of a packet that propagates through the communication system from the source to its destination. Since control usually implies a closed-loop setting, this delay gives a lower bound on how fast the controller may react to any system changes, thus limiting its effect on the system. Obviously this has a negative and even nonlinear effect on the control performance. 

The most intuitive and typical constraint is that the worst case delay has to be smaller than the time interval with which the controller operates. If so,  the controller does not experience any delay. A small worst case delay potentially allows for greater clock rates, leading to higher control performance. While there exists a good understanding of how such delay influences the control system \citep{memb_optimal_control_under_delay}
\citep{walsh_satbility_of_ncs}, only few ideas have emerged that go beyond such simple models. In 
\citep{goswami_codesign_cps}, a co-design of communication and control is proposed that allows for a distributional relaxation of delay constraints, such that the worst case delay may be larger than the controllers' time interval of operation. Another currently investigated idea evolves around event-based control schemes, helping to lower the requirements on the communication system while maintaining control performance 
\citep{linsenmayer_event_based_vehicle}, \citep{gross_event_based}. Additionally, a protocol design, specifically tailored to the control needs, can lower effects of jitter and packet loss \citep{mager_wireless_feedback}, leading also to delay minimization.

From the control perspective, achieving common control objectives for various autonomous and dynamically (de-) coupled subsystems is a challenge, in particular if input and coupled or uncoupled state constraints have to be satisfied. In \citep{dunbar2006distributed}, an approach of distributed model predictive control (DMPC) for this purpose was introduced, but communication of subsystems and delayed information exchange are neglected. In \citep{franco2008cooperative} and  \citep{gross2014distributed}, DMPC schemes are proposed by addressing constant communication delays. Time-varying delay is treated in \citep{zhang2016modified} according to a switching topology of feedback controllers, while neglecting input and state constraints. All these approaches aim at enabling a maximum control frequency for a given delay, or improving the control result under constant or worst case delay.

In this paper, we explore a different approach: The main idea is that the control algorithm takes \textit{reliable} packet delay predictions from the communication system, and gives this information to the corresponding plant controllers. Hence the controller is provided with the additional information, of when an expected set of data packets will arrive. The main contribution of this paper is to show: 1) how to create these delay predictions in the communication system and 2) how to use them in the control system to improve performance. Interestingly, to predict packet delay over some finite horizon is, to the best of our knowledge, a novel approach not been considered before.


\section{General Approach}

\label{sec:general_model}

\begin{figure}[b!]
\begin{center}
\psfrag{nr}{Communication Realm}
\psfrag{cr}[r][r]{Control Realm}
\psfrag{pla}[c][c]{Plant}
\psfrag{con}[c][c]{Controller}
\psfrag{cn}[c][c]{\small CN}
\psfrag{cn2}[c][c]{\tiny CN}
\psfrag{nc}[c][c]{Centralized Network Controller}
\psfrag{Communication Realm}{Communication Realm}
\includegraphics[width=7.6cm]{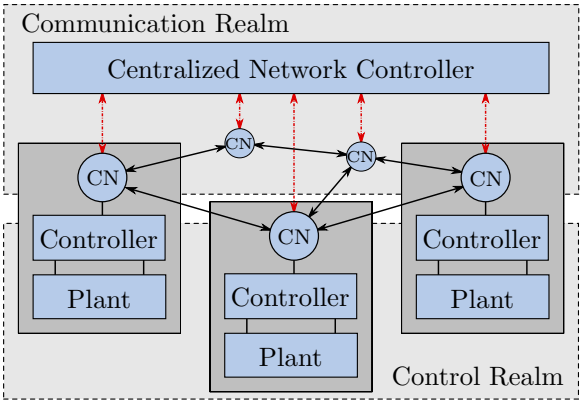}
\caption{\small General structure and communicated information. Each plant controller represents a node (CN) of the network.}
\label{fig::general_system_model}
\end{center}
\end{figure}

The entire system model consists of two major parts, see Fig. \ref{fig::general_system_model}: First, in a  distributed setting, multiple physical subsystems (plants) are each equipped with a local controller. These subsystems are assumed to be dynamically decoupled, but are coupled through a common control objective to perform a cooperative task. To achieve this objective, the plant controllers exchange information (like state and input trajectories) over a communication system subject to time-delay. Receiving a delayed state and input trajectory, an uncertain nominal state trajectory of the sending subsystem can be reconstructed by knowing the plant dynamics of the sender. Having an upper bound of possible uncertainty caused by communication delay and of possible deviations from previous communicated predictions, each subsystem can use robust MPC (RMPC) with respect to these uncertainties. By repeatedly solving local RMPC optimization problems for each subsystem, the common control goal can be achieved.

The underlying communication system itself consists of multiple communication nodes (CN), which are not necessarily all located at the local controllers. The data is routed over the communication system, and the routing is controlled by a dedicated, centralized network controller. The functionality of the overall model is as follows:
\begin{enumerate}
\item The network controller not only routes and schedules the data flows generated by the plant controllers over multiple hops in the network, but is designed to also predict packet delays over a future horizon.
\item This prediction contains information about when data will arrive at the plant controllers with (to a certain degree adjustable) reliability. Reliability is achieved through simple repetition of sending a packet in a hop. This leads to so-called \textit{delay trajectories}, which are communicated in each time instance to the plant controllers (red arrows in Fig. \ref{fig::general_system_model}).
\item The plant controllers are designed to exploit these \textit{delay trajectories} locally by RMPC to achieve an improved common control performance.
\end{enumerate}

To exchange data, each controller acts as a communication node (in a possibly larger network). The dedicated network controller has knowledge of all nodes with their states, of the momentary network topology (meaning accessible communication links between the nodes), and of where the data is scheduled to arrive. Furthermore, the network controller is implemented as an MPC, too. Only then, we are able to yield future communication schedules, which are necessary for predicting delay times of exchanged data by the plant controllers. Note that within this scheme, we assume that transmission along the communication network is time-consuming, however the network controller can send its predicted delay trajectories as well as its control inputs without any delay directly to the communication nodes and associated plant controllers. This is a valid assumption as long as the size of data, exchanged by the plant controllers, is considerably larger then the size of the delay trajectories.

The next two sections will go into the details of the communication and control system design.

\section{Communication System}

\label{sec:com_sys}

We model the communication network with a discrete-time, packet-based queueing system. Herein, each network entity $i$ is assigned to one queue $q^i$, which is $1$, if a corresponding packet is present at the entity, and $0$ otherwise. Hence, packet transmission can be modeled as increasing (and decreasing) queues by 1.
We denote all queues as a \textit{queue vector} $q = (q^1\dots q^{n_q})^T \in \{0,1\}^{n_q}$, where $n_q$ is the number of entities in the system. If entity $i$ wants its information to be transported to some destination $j$ (a \textit{communication request}), the queuing system is initialized with $q^i$ being $1$ while all other entries in $q$ are 0.
The system evolves with $k$ according to:
\begin{equation}
\label{eq::simple_system_evolution}
q_{k+1} = q_k + R_k v_k.
\end{equation}
Entities can exchange packets if there exists at least one communication \textit{link} between them. A link is a vector, which adds 1 to a queue (and possible subtracts 1 from another). They are collected as columns in the routing matrix $R \in \{ -1 , 0 , 1 \}^{n_q \times n_v}$ and can be activated by a network controller through a binary control vector $v_k \in \{ 0,1 \}^{n_v}$. Due to wireless effects, transmission over links may only succeed with probability $p^j_k$ (link $j$, time $k$). Therefore, the system evolves with $R_k = R \cdot \operatorname{diag}_{j} \{ \mathbb{B}[ p^j_k] \}$, where $\mathbb{B}[\cdot]$ is a Bernoulli trial. Note that any controller only knows $R$ and $p^j_k$ but \textit{not} $\mathbb{B}[p^j_k]$; specific modeling of the process $R_k$ can be found in \citep{pnc}. Furthermore, we usually cannot activate all links at the same time, expressed by the so-called constituency matrix $C\in \{ 0,1 \}^{n_c\times n_v}$ and the constraint $C v_k \leq \ONE_{n_c}$, where $\ONE_{n_c}$ is a vector of ones with dimension $n_c$. 
Finally, let $q^i$ be the queue of the entity, that the packet is destined for, then the goal of the network controller is to find a sequence of activations $v_k$ to make $q^i = 1$.

The system so far only models transportation of one data packet and is thus referred to as a subsystem. To trace transportation of several data packets (with possibly differing origin and destination), we initialize a new copy of the subsystem each time, a communication request occurs. We can stack these subsystems together in a block diagonal manner since they will only be coupled through the constituency constraints. With slight abuse of notation we will remain with the introduced notation, however from now on are extending its meaning to include stacked variables. Note that each time, a communication request has been served, meaning that the packet has arrived at its destination, we can remove the subsystem from the stack.

Let us define a suitable control policy, that we call \textit{Reliable Predictive Network Control} (R-PNC). The goal of this policy is not only to transport data packets but also to deliver a \textit{reliable delay forecast} over a time horizon $H$ in each time step $k$, providing information of when data will arrive at its destination. To achieve this, we define a scheduled transmission to be \textit{reliable}, if its overall transmission success probability, determined through $p^j_k$, is greater than some threshold $\phi$. Consequently, if a link has a small instantaneous success probability it has to be activated multiple times consecutively to increase the overall transmission success probability. In detail, we define $r_{k}^j \in \mathbb{N}$, for each link $j=1,\dots,n_v$, counting the amount of repetitions needed to ensure reliable communication for that link, when first activated in time step $k$.
If all $p_k^j$ are governed by a Discrete Time Markov Chain, calculating $r_k^j$ is a straight forward task for the whole prediction horizon. 

\begin{figure}
\begin{center}
\psfrag{CN1}[c][c]{\scriptsize $\text{CN}_1$}
\psfrag{CN2}[c][c]{\scriptsize $\text{CN}_2$}
\psfrag{CN3}[c][c]{\scriptsize $\text{CN}_3$}
\psfrag{2,1,2,...}[c][c]{${\color{red}2},2,1,\dots$}
\psfrag{3,1,1,...}[c][c]{$3,1,{\color{red}1},\dots$}
\psfrag{4,2,3,...}[c][c]{$4,3,2,\dots$}
\psfrag{rjjrjjrjj}[c][c]{$r^j_0,r^j_1,r^j_2,\dots$}
\includegraphics[width=0.325\textwidth]{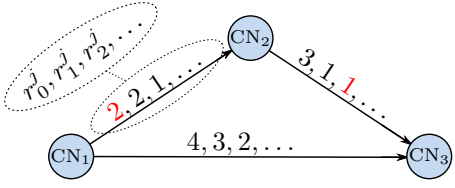}    
\caption{\small Graph Model with weight sequences: shortest path from $\text{CN}_1$ to $\text{CN}_3$ goes through $\text{CN}_2$} 
\label{fig::sequenced_graph}
\end{center}
\end{figure}

With this, a weighted graph can be constructed, in which nodes represent communication nodes, and edges represent the communication links. In this graph, $r_k^j$ are sequences (in $k$) of weights for each edge $j$, representing how many time steps one \textit{has to} spend repeating that edge, when starting at time $k$, before reaching the next node.
For a reliable prediction of communication delays between two plant controllers, the network controller has to determine a path between the corresponding nodes. The elapsed time for that path  is given by the sum of the specific entries of the weight sequences along the way as shown in Fig. \ref{fig::sequenced_graph}. Here we identify the shortest path (in red numbers) by $\tau_k\s{\text{CN}_1,\text{CN}_3}=3$, so an information send from CN$_1$ at time $k$ arrives at time $k+3$ at CN$_3$. These delay times are used in Sec. 4 by the control system.
Note that due to the definition of $r_k^j$, the real communication will probably be much faster then the predicted one. To consider the $r^j_k$ in the system evolution, we define:
\begin{equation}
\Gamma_\kappa = \underset{k=0,\dots, H-1}{\text{diag}} \left \{ \underset{j = 1, \dots, n_v}{\text{diag}} \left \{ \Theta \left[ \kappa - r_k^j - k \right] \right \} \right \}
\end{equation}
where $\text{diag} \{ \cdot \}$ is the diagonal matrix of its arguments, and $\Theta[ \cdot ]$ is the Heaviside function. The matrix $\Gamma_\kappa \in \mathbb{N}^{H \cdot n_v \times 
H \cdot n_v}$ ($\kappa$ being the time index for the prediction) transforms $r_k^j$ into a mask function for the control vector, being $0$, when the necessary amount of repetitions for reliability has not been reached yet, and $1$ otherwise.

The algorithm finds the \textit{best} path by minimizing a cost function $J_H$, which assigns costs to each trajectory of queue states up until the prediction horizon $H$. The cost of a queue state can be defined in a quadratic fashion through a cost matrix $Q_q$, such that:
\begin{equation}
\label{eq::algo_cost_function}
J_H(q_0) = \sum_{l=1}^H q_l^T Q_q q_l
\end{equation}
We set the cost of the queue, that represents the destination entity, to a global minimum of $q_H^T Q_q q_H$. Note that for the algorithm to work, $H$ has to be larger then the fastest weighted way between origin and destination.

\setlist[itemize]{wide=\parindent}

For simplicity of notation, we will now assume that the current time step is $k= 0$.
The predicted future queue state is determined by the planned control decisions. We collect all planned control decisions over the horizon $H$ in a control trajectory
$\tr{v}_0^T = \begin{pmatrix}
v_0^T, & v_1^T, \dots, & v_{H-1}^T
\end{pmatrix}$.
Through their definition, $r_k^j$ and $\Gamma_\tau$ contain the processed information of the stochastics of the communication model. Knowing $\Gamma_\tau$, the algorithm can therefore use a deterministic system evolution for its prediction, which will hold in a worst case sense:
\begin{equation}
\label{eq::algo_system_evolution}
q_\kappa = q_0 + \left[ \ONE^T_H \otimes R \right] \Gamma_\kappa \tr{v}_0 \ , \qquad \kappa = 1, \dots, H.
\end{equation}
Additionally, the system has to abide by the following constraints:
\begin{itemize}
\item \textit{Constituency Constraints} represent disjunct control decisions, e.g. due to physical limitations, and can be expressed via:
\begin{equation}
\label{eq::algo_c1}
\left[ I_H \otimes C \right] \tr{v}_0 \leq \ONE_{Hn_c}
\end{equation}
\item \textit{Reliability Constraints} force the controller to consider the necessary repetitions of scheduled control decisions, in order to guarantee reliable communication. 
More specifically, these constraints forbid the controller to do any disjunct control decision for the appropriate amount of time steps.
Defining $c^i_j$ as a row of $C$, which is $1$ at the $j$-th entry, $e_j$ as the canonical unit vector of dimension $n_v$, $\ZERO$ as the zero vector of dimension $n_v$. 
Then for every $j = 1, \dots, n_v$ and  $\kappa = 0 ,\dots, H-1 $ and every possible $i$, we have to construct the constraint:
\begin{equation}
\label{eq::algo_c2}
\begin{pmatrix}
\left[ \ONE_\kappa^T \otimes \mathbf{0}^T \right] 
& 
e_i^T
& 
\left[ \ONE^T_{r_\kappa^j-1} \otimes c_j^k \right] 
&
\left[ \ONE_{H-\kappa-r_\kappa^j}^T \otimes \mathbf{0}^T \right]
\end{pmatrix} \tr{v}_0 \leq 1
\end{equation}
\item \textit{Consistency Constraints} guarantee, that a once communicated arrival time can \textit{at worst} stay the same in a future optimization. Let $\delta(i)$ be the index of the destination queue of data packet $i$, $a(i)$ the arrival time (relative to the time of optimization) of that packet at that queue, and $\Delta k$ the elapsed time between the last optimization and now. Then it must hold for each possible data packet $i$ that:
\begin{equation}
\label{eq::algo_c3}
-
e^T_{\delta(i)} \left[ \ONE_H^T \otimes R \right] \Gamma_{ a(i) - \Delta k }
 \, \tr{v}_0
 \leq -1 +
e^T_{\delta(i)} q_0,
\end{equation}
where this time, the dimension of $e_i$ is $n_q$.
\item \textit{Processability Constraints} ensure that all queues are positive or zero at all times and forbid the routing of the same data through multiple nodes in one single time step. Defining $R^-$ as the copy of routing matrix $R$ in which all positive entries are set to $0$, and $R^+$ in the reverse way, we have to implement for all $\kappa = 1 , \dots, H$:
\begin{equation}
\label{eq::algo_c4}
-\left[
\begin{pmatrix}
\ONE_{\kappa}^T \otimes R^-
&
\ONE_{H-\kappa}^T \otimes \ZERO
\end{pmatrix} 
+
\left[ \ONE_H^T \otimes R^+ \right] \Gamma_{ \kappa -1}
\right]
\tr{v}_0
\leq q_0
\end{equation}
\end{itemize}

This completes the constrained optimization problem of minimizing \eqref{eq::algo_cost_function}. Note that the processability constraints can be relaxed to fit the physical situation. Furthermore, the consistency constraints have to be deactivated in the rare event, that reliable communication does not succeed (w.p. $1-\phi$).

Finally, the network control policy steers the network according to the following scheme:
\begin{enumerate}
\item For the stacked system, minimize \eqref{eq::algo_cost_function} following the system evolution \eqref{eq::algo_system_evolution} subject to \eqref{eq::algo_c1}, \eqref{eq::algo_c2}, \eqref{eq::algo_c3}, and \eqref{eq::algo_c4}.
\item Calculate the expected delay trajectories, and communicate them to the plant controllers.
\item Apply the first part of the optimal control vector trajectory $\tr{v}^*_0$.
\item Recognize the new system state and the newly requested data transmission, delete and add subsystems accordingly.
\item Repeat.
\end{enumerate}

\section{Control System}
\label{sec:con_sys}
Consider a plant that is partitioned into a set of subsystems, each modeled as discrete-time LTI system:
\begin{equation}\label{eq:dyn}
 x\s{i}_{k+1} = A\s{i} x_k\s{i} +B\s{i} u_k\s{i},
\end{equation}
where $i\in \mathcal{N}$ indicates the subsystem, $x_k\s{i} \in \mathbb{R}^{n_x\s{i}}$ the local state, and $u_k\s{i} \in \mathbb{R}^{n_u\s{i}}$ the local control input. The state and input are subject to polytopic constraints.
While the subsystem dynamics \eqref{eq:dyn} itself are uncoupled, we consider the case that a common control goal has to be reached in the sense that a subsystem $i\in \mathcal{N}$ minimizes a cost function which depends on the state and/or input of other subsystems $j\neq i$. In order to model from where information is required, a
predecessor set $\mathcal{N}_p\s{i}$ is defined, which contains the indices of those subsystems $j$ from which subsystem $i$ receives information. Likewise $\mathcal{N}_f\s{j}$ denotes the follower set containing the indices of subsystems $i\neq j$ which receive information from $j$.
\begin{assum}
Is is assumed that the information structure established by the sets $\mathcal{N}_p\s{j}$ and $\mathcal{N}_f\s{j}$ is acyclic. Furthermore, we assume that the communication between two subsystems ($j$: sender, $i$: receiver) is subject to the time-varying communication delay \rev{  $\tau_k\s{j,i}$ introduced in Sec. 3. Additionally, we introduce $d_k\s{i,j}$ as the age of the newest information $i$ has at time step $k$ of subsystem $j$}.
\end{assum}
Given the PNC-policies introduced in Sec.~\ref{sec:com_sys} the following holds: If $i \in  \mathcal{N}_f\s{j}$, then subsystem $j$ knows the time-delay \rev{$\tau_k\s{j,i}$}, and if $j \in \mathcal{N}_p\s{i}$, then subsystem $i$ knows the \rev{age $d_{k}\s{i,j}$} of the next incoming information sent by $j$. Note that in general \rev{ $\tau_{k}\s{j,i} \neq d_{k+\tau_k\s{j,i}}\s{i,j}$} (since newer information with small delay may \textit{overtake} older information).

Now, let $u_{k+l|k}\s{i}$ denote the prediction of subsystem $i$ of its local input at time $k+l$, calculated and sent at time $k$. Furthermore, let $u_{k+l|k}\s{i,j}$ for $j\in\mathcal{N}_p\s{i}$ be the prediction that subsystem $i$ has of the inputs $u_{k+l|k}\s{j}$ of subsystem $j$.
Since there is a communication delay and subsystem $j$ is, at time $k+l$, not restricted to choose $u_{k+l}\s{j}$ equal to the value $u_{k+l|k}\s{i,j}$ as predicted and communicated earlier, subsystem $i$ needs to consider this possible deviation. It is denoted by $\delta u_{k+l|k}\s{i,j}$, and subsystem $i$ considers this uncertainty by using a variable $\overline{u}_{k+l|k}\s{i,j}=u_{k+l|k}\s{i,j}+\delta u_{k+l|k}\s{i,j}$ when considering the dynamics of $j$ for evaluating its cost function.

In addition to the inputs, subsystem $j$ communicates its current state $x_{k|k}\s{j}$ at time $k$. The subsystem $i$ knows this state exactly at time-step \rev{$k'=k+\tau_k\s{j,i}$}. Thus, the last state of subsystem $j$ that is exactly known to $i$ is \rev{$x_{k-d_{k}\s{i,j}|k-d_{k}\s{i,j}}\s{i,j}$}. With this last exactly known state and with the predicted input trajectory, subsystem $i$ can estimate the state of $j$:
\vspace*{-1.5mm}
\begin{equation}\rev{
x_{k|k}\s{i,j}= A\s[d_{k}\s{i,j}]{j}x\s{i,j}_{k-d_{k}\s{i,j}|k-d_{k}\s{i,j}}+\sum_{l=1}^{d_{k}\s{i,j}}A\s[l-1]{j}B\s{j}u_{k-l|k}\s{i,j}.}
\end{equation}
Using the uncertain input and the state estimation, an augmented prediction model can be formulated. Assuming for simplicity of notation that subsystem $i$ has just one predecessor $j$, this model is:
\begin{equation}
\mathbf{x}_{k+l+1|k}\s{i} = \mathbf{A}\s{i} \mathbf{x}_{k+l|k}\s{i}+ \mathbf{B}\s{i}u_{k+l|k}\s{i}+ \mathbf{B}_1\s{i} \left( u_{k+l|k}\s{i,j}+ \delta u_{k+l|k}\s{i,j} \right),
\label{eq::augmodel}
\end{equation}
with the state vector $\mathbf{x}_{k+l|k}\s{i}= \begin{bmatrix}
x_{k+l|k}\s[T]{i}, x_{k+l|k}\s[T]{i,j}\end{bmatrix}^{T}\in \mathbb{R}^{\textbf{n}_x\s{i}}$, the vector of input uncertainties of the predecessor $\delta u_{k+l|k}\s{i,j}\in\mathbb{R}^{n_u\s{i}}$ where $\textbf{n}_x\s{i} = \sum_{j\in \mathcal{N}_p\s{i}\cup\{i\}}n_x\s{j}$ as well as:
\begin{align}
\mathbf{A}\s{i} =\begin{bmatrix}
A\s{i} & 0\\
0 & A\s{j}
\end{bmatrix},\
\mathbf{B}\s{i} =\begin{bmatrix}
B\s{i}\\
0\\
\end{bmatrix},\ 
\mathbf{B}_1\s{i}=\begin{bmatrix}
0 \\
B\s{j}
\end{bmatrix}
\end{align}
The stacked state vector $\mathbf{x}_k\s{i}$ is subject to a polytopic constraint $\mathbb{X}\s{i} =\{ \mathbf{x}_k \big\vert \mathbf{C}_x\s{i} \mathbf{x}_k \leq \mathbf{b}_x\s{i} \}$.
 
Each subsystem can now use such a prediction model within a robust MPC scheme to determine its own control inputs. The goal is to guarantee robustness with respect to the defined uncertainties arising from communication delay and the deviations from previously communicated trajectories. Thereto, the disturbance-feedback policy proposed in 
\citep{gross2014distributed} can be enhanced by the predicted time-delay of the next incoming information \rev{$d_{k+l}\s{i,j}$} to the following representation:
\vspace*{-1.5mm}
\begin{equation}
u_{k+l|k}\s{i}=v_{k+l|k}\s{i}+\sum_{r=1-\rev{d_{k}\s{i,j}}}^{l-\rev{d_{k+l}\s{i,j}}} K_{l,r|k}\s{i} \delta u_{k+r|k}\s{i,j},\label{eq::dfb}
\end{equation}
where $v_{k+l|k}\s{i}$ is the control input in absence of uncertainties, and $K_{l,r|k}\s{i}$ is the feedback-gain to account for the uncertainties. By formulating stacked vectors over the prediction horizon $H$ for state,  inputs, and uncertainties:
\begin{flalign}
\tbf{x}_k\s{i} &= {[\mathbf{x}_{k|k}\s[T]{i}, \dots,\mathbf{x}_{k+H|k}\s[T]{i}]}^T,\;\tilde{u}_k\s{i} = {[u_{k|k}\s[T]{i}, \dots,u_{k+H-1|k}\s[T]{i}]}^T
\nonumber\\
\tilde{u}_k\s{i,j} &= {[u_{k|k}\s[T]{i,j},\dots,u_{k+H-1|k}\s[T]{i,j}]}^T,\nonumber\\
\delta\tilde{u}_k\s{i,j} &={[\delta u_{k+1-\rev{d_{k}\s{i,j}}|k}\s[T]{i,j},\dots,\delta u_{k+H-1|k}\s[T]{i,j} ]}^T,  \end{flalign}
the following representation is obtained:
\begin{equation}
\tbf{x}_k\s{i} = \tbf{A}\s{i}\textbf{x}_{k|k}\s{i}+\tbf{B}\s{i}\tilde{u}_k\s{i}+\tbf{B}_1\s{i}  \tilde{u}_k\s{i,j} + \tbf{B}_2\s{i} \delta\tilde{u}_k\s{i,j}.
\label{eq::predmodel}
\end{equation}
Equation \eqref{eq::dfb} can be rewritten to:
\begin{equation}
\tilde{u}_k\s{i} =\tilde{v}_k\s{i}+\tilde{K}_k\s{i} \delta\tilde{u}_k\s{i,j},
\label{eq::dfb_big}
\end{equation}
with $ \tilde{v}_k\s{i} \in \mathbb{R}^{H\,n_u\s{i}}$ and the uncertainty feedback matrix $\tilde{K}_k\s{i} \in \mathbb{R}^{H\,n_u\s{i}\times n_{1,k}\s{i}}$, where $n_{1,k}\s{i}\left(\rev{d_{k}\s{i,j}}\right)$ represents the time varying length of $\delta\tilde{u}_k\s{i,j}$. Similarly as in \citep{goulart2006optimization}, the lower triangular block matrix $\tilde{K}_k$ can be enhanced with the prediction of  \rev{$\tilde{d}_{k}\s{i,j}=[d_{k}\s{i,j}, \dots,d_{k+N-1}\s{i,j}]$}.

Let constraints for the stacked vectors of state, input, and uncertainties over the prediction horizon be denoted by:
\begin{flalign}
\tbf{x}_k \in\tilde{\mathbb{X}}\s{i},\quad \tilde{u}_k\in 
\tilde{\mathbb{U}}_k\s{i},\quad \delta\tilde{u}_k\in \delta\tilde{\mathbb{U}}_k\s{i}.
\label{eq::Constraints}
\end{flalign}
When introducing auxiliary matrices $\mathbf{F}_1$ to $\mathbf{F}_5$ according to \citep{gross2014distributed}), the admissible set of input sequences can be defined to:
\begin{equation}
	\Pi_k\s{i}\left(\mathbf{x}_{0}\right) = \left\{ \left(\tilde{K}_k\s{i},\,\tilde{v}_k\s{i}\right) \left\vert \begin{array}{l}
	\mathbf{x}_{k|k}\s{i}=\mathbf{x}_0,\; \exists Z_k\s{i}\geq 0:\\
	 Z_k\s{i}\tbf{C}_\delta\s{i}=\mathbf{F}_2\s{i}\tilde{K}_k\s{i}+\mathbf{F}_4\s{i}\\
	\mathbf{F}_2\s{i}\tilde{v}_k\s{i}+Z_k\s{i}\tilde{b}_{\delta_k}\s{i} \leq \dots \\
	\quad \mathbf{F}_5\s{i}-\mathbf{F}_1\s{i}\mathbf{x}_{k|k}\s{i}-\mathbf{F}_3\s{i}\tilde{u}_k\s{i,j}
\end{array}	  \right.\right\}, \label{eq::aSis2}
\end{equation}
with slack variables $Z_k\s{i}=\left[Z_{x_{k|k}}\s[T]{i}, \; Z_{u_{k|k}}\s[T]{i} \right]^T$, allowing us to characterize the 
set of possible reactions to uncertainties in \eqref{eq::dfb_big} by:
\begin{equation}
\Delta\tilde{\mathbb{U}}_k\s{i} = \left\{\Delta\tilde{u}_k\s{i} \in\mathbb{R}^{H\,n_u\s{i}} \left\vert \tilde{C}_u\s{i}\; \Delta\tilde{u}_k\s{i} \leq Z_{u_{k|k}}\s{i} \tilde{b}_{\delta_k}\s{i}=\tilde{b}_{\Delta_k}\s{i} \right.\right\}, \nonumber
\end{equation} 
with $\Delta\tilde{u}_k\s{i}= \tilde{K}_k\s{i}\delta\tilde{u}_k\s{i,j}$ and $\tilde{b}_{\Delta_k}\s{i}=\begin{bmatrix}
b_{\Delta_{k|k}}\s[T]{i},\dots,b_{\Delta_{k+H-1|k}}\s[T]{i}
\end{bmatrix}^T$.
This set represents also the set of uncertainties to be communicated to a successor of subsystem $i$.
With respect to terminal sets and terminal control laws to establish recursive feasibility, we refer the reader to the solution proposed in \citep{gross2014distributed}.

For each subsystem, a two-stage optimization problem is now to be solved in any $k$: The first stage minimizes the cost function formulated to be quadratic in the augmented state and control inputs (thus depending on the planned and communicated inputs of the predecessors), but without considering the uncertainties:
\vspace{-1mm}
\begin{align}
&J_1\s{i} = \norm{\mathbf{x}_{k+H|k}\s{i}}^2_{Q_T\s{i}} +\sum_{l=0}^{H-1} \norm{\mathbf{x}_{k+l|k}\s{i}}^2_{Q_x\s{i}}+\norm{\begin{bmatrix}
u_{k+l|k}\s[T]{i} & \tilde{u}_{k+l|k}\s[T]{i,j}
\end{bmatrix}}^2_{Q_u\s{i}} \nonumber\\
& V_1\s{i}\left(\mathbf{x}_0\s{i}\right) := \min_{\tilde{v}_k\s{i},\,\tilde{K}_k\s{i}} J_1\s{i} \label{eq::V1}\\
&\text{s.t.: } \mathbf{x}_{k|k}\s{i}= \mathbf{x}_0\s{i}, \quad 
	 \left(\tilde{K}_k\s{i} ,\, \tilde{v}_k\s{i} \right)\in\Pi_k\s{i}\left(\mathbf{x}_0\s{i}\right),\eqref{eq::Constraints}, \eqref{eq::aSis2},\nonumber\\ 
& \qquad  \eqref{eq::augmodel},\eqref{eq::dfb} \text{ with } \delta u_{k+l|k}\s{i,j} =0\ \forall l \in \{0,\dots,H{\small -}1\} \nonumber.
\end{align}
The entries in $\tilde{K}_k\s{i}$ are not uniquely defined for $V_1\s{i}\left(\mathbf{x}_0\s{i}\right)$, and since $\tilde{K}_k\s{i}$ affects the set $\Delta\tilde{\mathbb{U}}_k\s{i}$ , $\tilde{K}_k\s{i}$ is optimized in a second stage, still satisfying \eqref{eq::aSis2}. Using two uncertainty sets $\delta\tilde{\mathbb{U}}_k\s{i} = \begin{bmatrix}
\delta\mathbb{U}_{k+1-d_{k}\s{i,j}|k}\s{i}, \dots, \delta\mathbb{U}_{k+H-1|k}\s{i}
\end{bmatrix}$ and $\Delta\tilde{\mathbb{U}}_k\s{i} = \begin{bmatrix}
\Delta\mathbb{U}_{k|k}\s{i}, \dots, \Delta\mathbb{U}_{k+H-1|k}\s{i}
\end{bmatrix}$, this cost function is defined as:
\vspace{-1.5mm}
\begin{equation}
J_2\s{i}=\sum_{l=1}^{H-\rev{d_{k}\s{i,j}}-1} f_1\left( \delta\mathbb{U}_{k+l|k}\s{i}, \Delta\mathbb{U}_{k+l|k}\s{i}\right)+ \sum_{l=0}^{\rev{\tau_{k}\s{i}}} f_2\left(\Delta\mathbb{U}_{k+l|k}\s{i}\right), \nonumber
\end{equation}\\[-2mm]
where \rev{$\tau_{k}\s{i}= \max_{j\in\mathcal{N}_f\s{i}}\left(\tau_{k}\s{i,j}\right)$} is the maximum delay of outgoing information. In the cost function, $f_1$ is used to balance the incoming and outgoing uncertainty sets $\left(\delta\tilde{\mathbb{U}}_k\s{i},\Delta\tilde{\mathbb{U}}_k\s{i}\right)$, i.e.,  the main idea is to preserve for $k+l$ the same flexibility for adapting  $u_{k+l|k}\s{i}$ with $\Delta u_{k+l|k}\s{i} \in \Delta\mathbb{U}_{k+l|k}\s{i}$ as the predecessor has predicted for its control input through $\delta u_{k+l|k}\s{j}\in \delta\mathbb{U}_{k+l|k}\s{j}$.
This balancing avoids an blow-up of uncertainty sets by propagating the uncertainties through the interconnected graph. The term $f_2$, on the other hand, is used to tighten the uncertainty set $\Delta\tilde{\mathbb{U}}_k\s{i}$ for the next \rev{$\tau_k\s{i}$} time steps required to pass information to the follower.

The optimization of the second stage is now:
\begin{flalign}
V_2\s{i}\left(\mathbf{x}_0\s{i}, \tilde{v}_k\s{i}\right) &= \min_{\tilde{K}_k\s{i}} J_2\s{i}\label{eq::V2}\quad\text{s.t.: } \left(\tilde{K}_k\s{i}  \right)\in\Pi_k\s{i}\left(\mathbf{x}_0\s{i},\, \tilde{v}_k\s{i}\right). 
\end{flalign}
%
Since the control input trajectory $\tilde{v}_k\s{i}$ is already computed in \eqref{eq::V1}, this is preset in \eqref{eq::V2} to the set of admissible input sequences $\Pi_k\s{i}$. Thus, \eqref{eq::V2} just adapts the values of $\tilde{K}_k\s{i}$ and, consequently, the behavior of the controlled system in the next time steps (but not in the current time step $k$).

\section{Simulation Example}
\label{sec:sim}
\begin{figure}[b!]
\psfrag{CN}[l][l]{Communication Realm}
\psfrag{CS}[l][l]{Control Realm}
\psfrag{CPS1}[c][c]{\small CPS 1}
\psfrag{CPS2}[c][c]{\small CPS 2}
\psfrag{CPS3}[c][c]{\small CPS 3}
\psfrag{c4}[c][c]{Centralized network controller}
\psfrag{obj}[c][c]{\tiny $CO$}
\psfrag{c1}[c][c]{\small Controller 1}
\psfrag{c2}[c][c]{\small Controller 2}
\psfrag{c3}[c][c]{\small Controller 3}
\psfrag{k1}[c][c]{\scriptsize CN$_1$}
\psfrag{k2}[c][c]{\scriptsize CN$_2$}
\psfrag{k3}[c][c]{\scriptsize CN$_3$}
\psfrag{k4}[c][c]{\tiny CN$_4$}
\psfrag{p1}[c][c]{\color{white} \small$\;$ Vehicle 1}
\psfrag{p2}[c][c]{\color{white} \small$\;$ Vehicle 2}
\psfrag{p3}[c][c]{\color{white} \small$\;$ Vehicle 3}
\includegraphics[width=8.8cm]{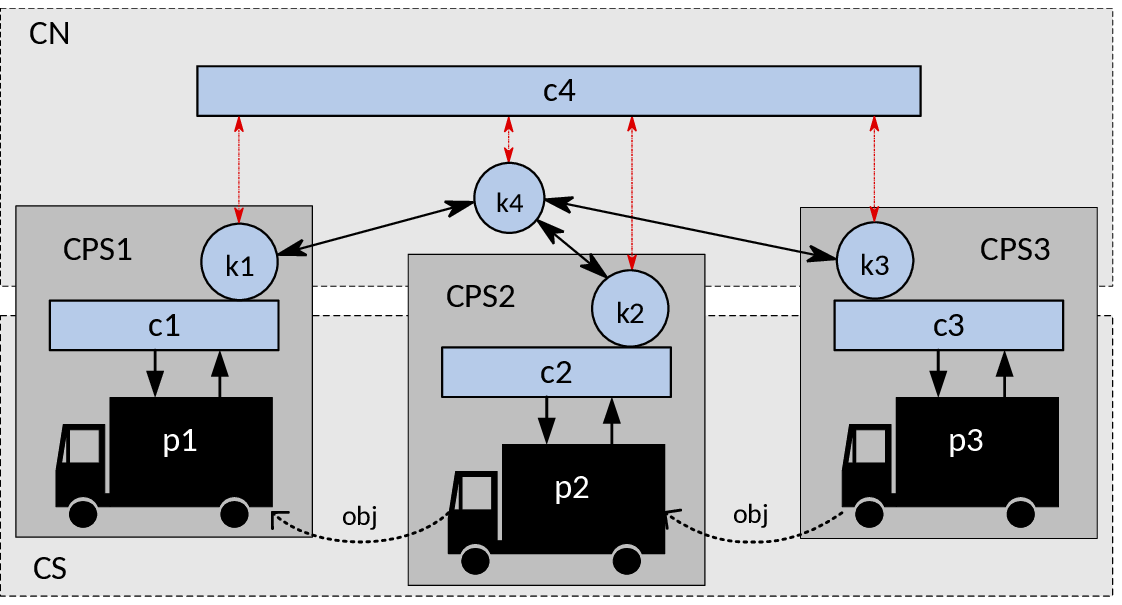}
\caption{\small Platooning example of a cyber-physical network. CN: Communication Node, CO: Control objective.}
\label{fig::CPN_example}
\end{figure}

\begin{figure}[t!]
\subfigure[CPS $1$]{\includegraphics[width=0.23\textwidth]{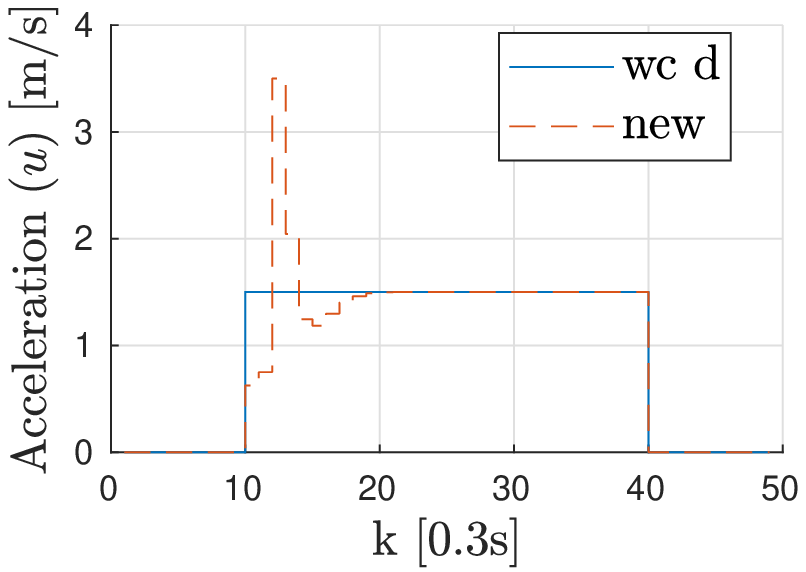}
\label{fig::cps1}}
\subfigure[CPS $2$]{\includegraphics[width=0.23\textwidth]{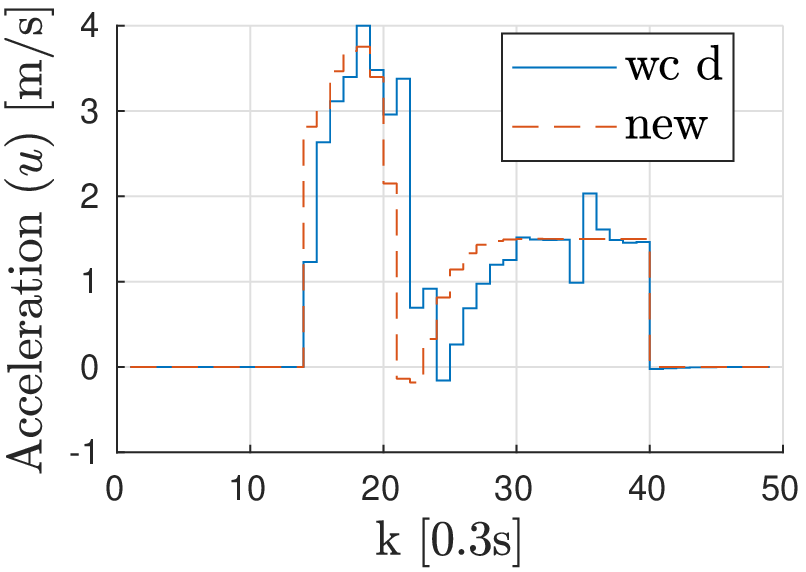}
\label{fig::cps2}}\\[-0.2cm]
\subfigure[CPS 1]{\includegraphics[width=0.23\textwidth]{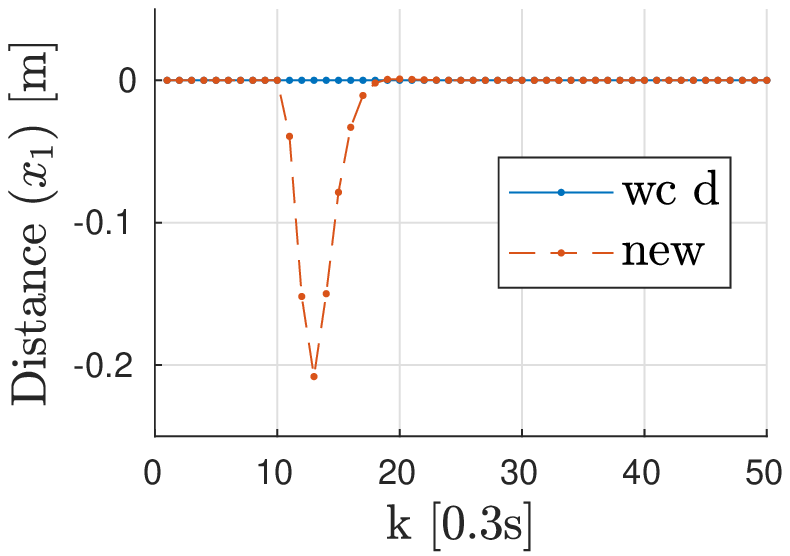}
\label{fig::cps3}}
\subfigure[CPS 2]{\includegraphics[width=0.23\textwidth]{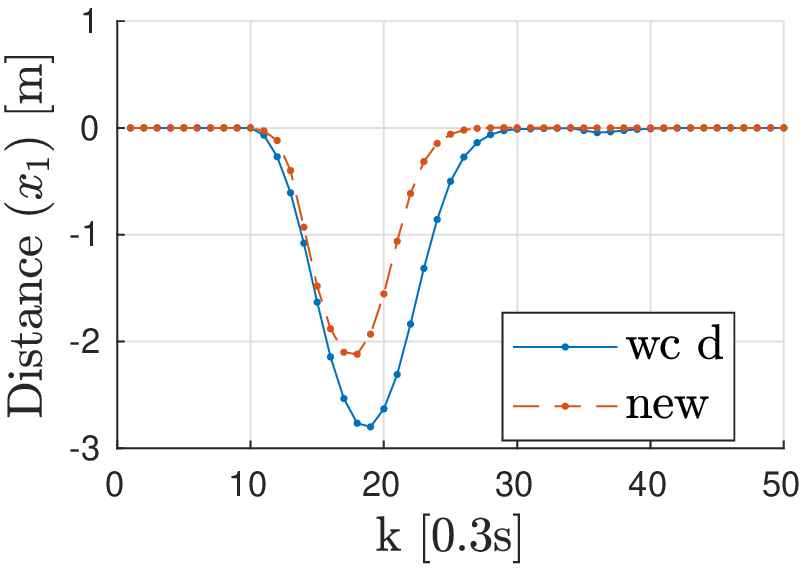}
\label{fig::cps4}}
\vspace{-0.4cm}
\caption{\small Control input and distance to reference}
\label{fig::cps}
\end{figure}

To illustrate the methods and the resulting gain in control performance, the approach is applied to the platooning example sketched in Fig.~\ref{fig::CPN_example}.
We assume that CPS 1 follows an uncertain trajectory, while CPS 2 and CPS 3 aim at maintaining the position and speed of the predecessor, while satisfying acceleration constraints.\\
The communication network is organized as a cellular network, so every incoming information is send to the base station CN$_4$, and then transmitted to a selected receiver. The selection and activation of communication links is controlled by the network controller with the introduced PNC-policies. For the sake of illustration, we restrict the delay \rev{\mbox{$\tau_k\s{2,3}=1$}} to a constant value. In contrast, \rev{{$\tau_k\s{1,2}$}} is time-varying, predictable by PNC-policies, and thus communicated to CPS 1 and CPS 2. Local dynamics and input constraints of all CPS are assumed to be identical:
\begin{equation}
x_{k+1}\s{i} = \begin{bmatrix}
1 & 0.3\\
0 & 1
\end{bmatrix} x_k\s{i}+\begin{bmatrix}
0.045\\
0.3
\end{bmatrix} u_k\s{i}, \quad u_k\s{i}\in\left[-4,\; 4\right]
\end{equation}
with $x_k\s{i}=\left[x_{1,k},\;x_{2,k}\right]^T$, where $x_{1,k}$ is the position, $x_{2,k}$ the velocity, and $u_k\s{i}$ the acceleration.
Hence, the augmented prediction model is given by $\mathbf{x}_{k}\s{i} = \left[ x_k\s[T]{i}, x_k\s[T]{i-1}\right]^T, i\in\{2,3\}$, and the augmented state constraints are specified as $\left( x_k\s{i}-x_k\s{i-1}\right)\in\left[10,\;10 \right]\times\left[10,\;10\right]$.
The cost function $J_2\s{i},i\in\mathcal{N}$ contains $f_1\left( \delta\mathbb{U}_{k+l|k}\s{i}, \Delta\mathbb{U}_{k+l|k}\s{i}\right)=\norm{b_{\delta_{k+l|k}}\s{i}-b_{\Delta_{k+l|k}}\s{i}}^2$, and $f_2\left(\Delta\mathbb{U}_{k+l|k}\s{i}\right)=\norm{b_{\Delta_{k+l|k}}\s{i}}^2$.
For a prediction horizon of $H=5$, it is assumed that the PNC-policies guarantee \rev{$\tau_k\s{1,2}\leq \bar{\tau}\s{1,2} = 4\ \forall k$, hence $d_k\s{2,1}\leq \bar{\tau}\s{1,2}\ \forall k$ holds}.
In the following, the simulation results of the presented methods (new) are summarized and compared to the worst case delay method (wc d) from \citep{gross2014distributed}.
The simulated scenario is the following: at \mbox{$k=0$}, the position, velocity, and acceleration of all vehicles are initialized to zero.
At $k=10$, the input reference trajectory of CPS 1 steps unpredicted to 1.5, thus CPS 1 starts accelerating while maintaining the previously communicated uncertainty sets.
At $k=40$, the reference steps back to zero, but this time predicted before. For the sake of comparison, the (in general time-varying) communication delay \rev{$\tau_{10}\s{1,2}$} is fixed at $k=10$ to \rev{the maximum value, so the step of CPS 1 is known to CPS 2 with $d_{14}\s{2,1}=4$}.
Fig.~\ref{fig::cps} shows the control input of CPS 1 (a) and CPS 2 (b). Since the new methodology already tightens the uncertainty set $\Delta\tilde{\mathbb{U}}_k\s{1}$ at $k=9$ while knowing the output delay, CPS 1 cannot react that agile at $k=10$ and $k=11$. Once the step of the control input is communicated, CPS 2 can react much more aggressive in $k=14$ and $k=15$ due to the smaller uncertainty set communicated by CPS 1 in $k=9$.
The slower behavior of the input of CPS 1 leads to a worse control result related to the reference, as plotted in Fig.~\ref{fig::cps3} (while CPS 1 follows the reference perfectly in the worst case delay simulation). This degradation is acceptable, and even intended, since CPS 2 can react much faster, as shown in Fig.~\ref{fig::cps4}.

Summing up all quadratic deviations between control result and reference trajectory according to the weights in $J_1$, the performance can be compared for both approaches, see Table \ref{tab::performance_measure}. The little worse result for CPS 1 (less than 2) has to be contrasted to drastic improvements for CPS 2 (more than 130 points) and even for CPS 3. Remembering that the communication delay \rev{$\tau_k\s{2,3} \equiv 1 $}, the better performance measure for CPS 3 is most significant. Since there is no difference between a worst case delay and a predicted delay of constant one, the improved control result is an effect of the second optimization stage.

\section{Conclusion}
\label{sec:conc}
We have proposed a new theory for combining predictive control of communication networks with distributed MPC for CPS. If a describing model of the communication network exists, it is advisable to control the communication network with the presented RPNC-algorithm. Reducing the overall buffer size, it produces a prediction of future transmitting delays, which is usable by the associated distributed control system in two ways: Firstly, incorporating the predicted age of future incoming data packages can increase the degree of freedom of computing the robust control invariant sets. Secondly, the prediction of sending delay can be used to optimize the control invariant set for the following CPS.
In comparison to a worst case communication delay, our methodology improves the overall control result of a platooning example by over $30\%$.

\begin{table}[t!]
\centering
\caption{\small Performance measure of platoon}
\vspace{-0.2cm}
\label{tab::performance_measure}
\begin{tabular}{|l||c|c|c|c|}
\hline
& CPS 1 & CPS 2 & CPS 3 & total\\
\hline\hline
worst case delay & 0 & 275.55 & 288.37 & 563.92\\
new methodology & 1.52 & 141.56 & 248.33 & 391.41\\
\hline
\end{tabular}
\end{table}

\bibliography{mpcpn_paper}             
                                                   







\end{document}